\documentclass[twocolumn,showpacs,preprintnumbers]{revtex4}
\usepackage{amssymb}
\usepackage{amsmath}
\usepackage{graphicx}
\usepackage{dcolumn}
\usepackage{bm}
\usepackage{color}

\begin{document}

\title{First-order and continuous quantum phase transitions in the
anisotropic quantum Rabi-Stark model}
\author{You-Fei Xie$^{1}$, Xiang-You Chen$^{1}$, Xiao-Fei Dong$^{1}$, and
Qing-Hu Chen$^{1,2,*}$}

\address{
$^{1}$ Zhejiang Province Key Laboratory of Quantum Technology and Device, Department of Physics, Zhejiang University, Hangzhou 310027, China \\
$^{2}$  Collaborative Innovation Center of Advanced Microstructures, Nanjing University, Nanjing 210093, China
 }\date{\today }

\begin{abstract}
Various quantum phase transitions in the anisotropic Rabi-Stark model with
both the nonlinear Stark coupling and the linear dipole coupling between a
two-level system and a single-mode cavity are studied in this work. The
first-order quantum phase transitions are detected by the level crossing of
the ground-state and the first-excited state with the help of the pole
structure of the transcendental functions derived by the Bogoliubov
operators approach. As the nonlinear Stark coupling is the same as the
cavity frequency, this model can be solved by mapping to an effective
quantum oscillator. All energy levels close at the critical coupling in this
case, indicating  continuous quantum phase transitions. The critical gap
exponent is independent of the anisotropy as long as the counter-rotating
wave coupling is present, but essentially changed if the counter-rotating
wave coupling disappears completely. It is suggested that the gapless
Goldstone mode excitations could appear above a critical coupling in the
present model in the rotating-wave approximation.
\end{abstract}

\maketitle

\section{Introduction}

The quantum Rabi model (QRM) describes the basic interaction between a
two-level (artificial) atom and a one-mode bosonic cavity ~\cite{Rabi,Braak2}
and is a paradigmatic model in quantum optics. In  conventional cavity
quantum electrodynamics (QED) systems, due to the extremely weak coupling
between the two-level systems and the cavity, the basic physics can be
explored in the rotating wave approximation (RWA) \cite{JC,book}. However,
the situation has changed in the past decade. In many advanced solid
devices, such as the superconducting circuit QED systems \cite%
{Niemczyk,Forn1} and trapped ions \cite{Leibfried,Clarke}, the ultrastrong
coupling even deep strong coupling \cite{Forn2,Yoshihara} between the
artificial atom and the resonators have been accessed, and the RWA is
demonstrated invalid \cite{Niemczyk}. On the other hand, the two-level
system appearing in these systems is just a qubit, which is the building
block of quantum information technologies with the ultimate goal being to realize
quantum algorithms and quantum computations. Just motivated by the
experimental advances and potential applications in quantum information
technologies, the QRM has attracted extensive attentions theoretically,
especially for the analytical solutions \cite{Casanova,chenqh,Braak,
Chen2012, Chen2,Zhong,zheng,luo2} and the  quantum phase
transition (QPT) ~\cite{plenio, hgluo}. For a  more complete review, please
refer to Refs. ~\cite{ReviewF,Kockum,Boit}.

The QRM continues to inspire exciting developments in both experiments and
theories recently. The anisotropic QRM~\cite{yejw2013,Fanheng,Tomka} was
motivated by the recent experimental progress \cite%
{Wallraff,Schiroa,Erlingsson}. It can be mapped onto the model describing a
two-dimensional electron gas with Rashba ( rotating wave coupling relevant)
and Dresselhaus ( counter rotating-wave coupling dependent) spin-orbit
couplings subject to a perpendicular magnetic field \cite{Erlingsson}. These
couplings can be tuned by an applied electric and magnetic field, allowing
the exploration of the whole parameter space of the model. This model can
directly emerge in both cavity QED \cite{Schiroa} and circuit QED \cite%
{Wallraff}. Interestingly, the first-order QPT is observed in the
anisotropic QRM ~\cite{Fanheng} and the Jaynes-Cummings model~\cite{puhan}.

On the other hand, Grimsmo and Parkins proposed a  scheme by adding a
nonlinear coupling term to the QRM Hamiltonian \cite{Grimsmo1,Grimsmo2}.
This nonlinear coupling term has been discussed in the quantum optics
literature under the name of dynamical Stark shift, a quantum version of the
Bloch-Siegert shift, so it was later named the quantum Rabi-Stark model
(RSM) \cite{Eckle}. This model has also attracted much attention in recent years
\cite{Maciejewski,Xie, Xie2,Cong}. More recently, the anisotropic Dicke
model with the Stark coupling terms, which can be called as anisotropic
Dicke-Stark model, was demonstrated via cavity assisted Raman transitions in
a configuration using counterpropagating laser beams~\cite{zhiqiang}. For the
one-atom case, it is just the anisotropic Rabi-Stark model (ARSM).

Actually, the implementation of the ARSM has been also demonstrated most
recently in a trapped ion~\cite{Cong}. In this proposal, external laser
beams are applied to induce an interaction between an electronic transition
and the motional degree of freedom, thus the Stark term is generated. The
anisotropic coupling strengths are determined by laser field amplitudes of
the red-sideband and blue-sideband driving lasers, which can be tuned
independently in the trapped ion experimental system.

Theoretically, the RSM has been studied by the Bargmann space approach~\cite%
{Eckle,Maciejewski}. Later it was solved by the Bogoliubov operator approach
(BOA) \cite{Xie}. Many exotic properties are found within the analytic exact
solutions, such as the first-order QPT and the spectra collapse \cite{Xie}.
Then what are the properties contained in the ARSM? Especially, since the
first-order QPT occur in the anisotropic QRM and the RSM, while the
continuous QPTs are present in the isotropic QRM, rich QPTs might appear in
this generalized model due to more tunable interaction constants in the wide
parameter space.

The paper is organized as follows. In Sec. II, we first describe the ARSM
briefly, and then demonstrate that eigensolutions can then be easily
obtained by the zeros of the transcendant function derived by the BOA. In Sec.
III, the first-order QPT is analyzed based on the level crossing of the
ground-state and the first-excited state, based on the pole structure of the
derived transcendant function. In Sec. IV, the energy gap near the critical
coupling for the nonlinear Stark coupling being the same as the cavity
frequency is calculated analytically by the exact mapping the ARSM to a
quantum oscillator. The energy-gap exponents are also obtained. The last
section contains some concluding remarks. Details of the solutions to 
the ARSM in the two cases are deferred to the Appendixes.

\section{Model and solutions}

The Hamiltonian of the ARSM reads
\begin{eqnarray}
H &=&\left( \frac{1}{2}\Delta +Ua^{\dagger }a\right) \sigma _{z}+\omega a^{\dagger
}a  \notag \\
&&+g_{1}\left( a^{\dagger }\sigma _{-}+a\sigma _{+}\right) +g_{2}\left(
a^{\dagger }\sigma _{+}+a\sigma _{-}\right) ,  \label{Hamiltonian}
\end{eqnarray}%
where $\Delta $ is qubit energy difference, $a^{\dagger }$ $\left( a\right) $
is the photonic creation (annihilation) operator of the single-mode cavity with frequency $\omega$, $%
g_{1}\ $and $g_{2}\ $ are the rotating-wave and counter rotating-wave
coupling constants, respectively, and $\sigma _{k}(k=x,y,z)$ are the Pauli
matrices. We define $r=g_{2}/g_{1}\ $as the anisotropic constant, which is
usually tuned by the input parameters. In this paper, the  unit is  set  $\omega =1$.

To explore the basic physics such as the various QPTs in this generalized
model, we will obtain the analytic exact solution by the BOA~\cite{Chen2012}%
. Associated with this Hamiltonian is the conserved parity $\Pi =\exp \left(
i\pi \widehat{N}\right) \ $where $\widehat{N}=\left( 1+\sigma _{z}\right)
/2+a^{\dagger }a$ is the total excitation number, such that $\left[ \Pi ,H%
\right] =0$. $\Pi $ has two eigenvalues $\pm 1$, depending on whether $%
\widehat{N}$ is even or odd. The parity symmetry not only facilitate to
study this model but also allow the possibility of the continuous QPT with
symmetry breaking.

Employing the following transformation
\begin{equation}
P=\frac{1}{\sqrt{2}}\left(
\begin{array}{ll}
\sqrt{r} & ~1 \\
-\sqrt{r} & \;1%
\end{array}%
\right) ,  \label{P1}
\end{equation}%
we have the transformed Hamiltonian $H_{1}=PHP^{-1}$ with the same
eigenenergy. Then we introduce two displaced bosonic operators with opposite
displacements.
\begin{equation}
A^{\dagger }=a^{\dagger }+w,B^{\dagger }=a^{\dagger }-w  \label{dis}
\end{equation}%
where $w$ is a displacement to be determined. The wavefunction can be
expanded in terms of the  $A$-operators%
\begin{equation}
\left\vert A\right\rangle =\left(
\begin{array}{c}
\sum_{n=0}^{\infty }\sqrt{n!}e_{n}|n\rangle _{A} \\
\sum_{n=0}^{\infty }\sqrt{n!}f_{n}|n\rangle _{A}%
\end{array}%
\right) .  \label{wave1}
\end{equation}%
where $e_{n}$ and $f_{n}$ are the expansion coefficients, $\left\vert
n\right\rangle_{A}$ is the bosonic number state in terms of the new photonic
operators $A^{\dagger }$ is
\begin{equation*}
\left\vert n\right\rangle_{A}=\frac{\left( A^{\dagger }\right) ^{n}}{\sqrt{n!%
}}D(-w)\left\vert 0\right\rangle ,
\end{equation*}%
where $D(w)=\exp \left( wa^{\dagger }-wa\right) $ is the unitary
displacement operator, $\left\vert 0\right\rangle $ is original vacuum state.

As described in detail in Appendix A, following the BOA, we can derive a
transcendant function to the ARSM, so called G-function
\begin{equation}
G_{\mp }\left( E\right) =\sum_{n=0}^{\infty }\left( e_{n}\pm f_{n}\right)
w^{n}=0,  \label{G-Func}
\end{equation}%
where $e_{n}$ and $f_{n}$ can be obtained from $f_{0}=1$ recursively in Eqs.
(\ref{relation1}) and (\ref{relation2}), $\mp $ corresponds to odd(even)
parity. According to Eq. (\ref{shift_ARS}), the G-function is well defined
in the regime $\left\vert U\right\vert <1$

\begin{figure}[tbp]
\includegraphics[width=\linewidth]{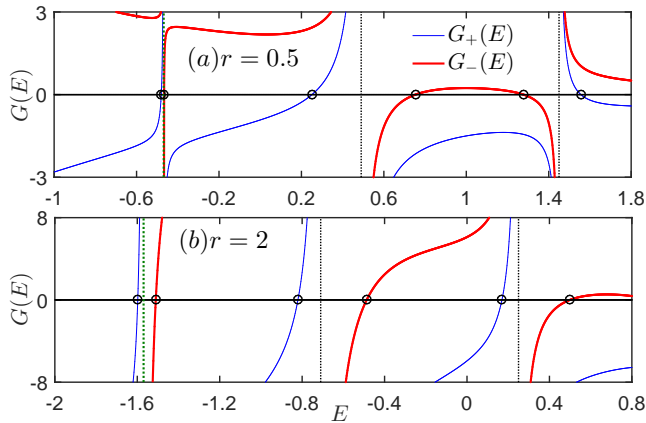}
\caption{ (Color online) G-curves for $\Delta =0.7,g_{1}=0.8,U= 0.2$,$r=0.5$
in the upper panel and $r=2$ in the lower panel. Thin  blue lines and thick red
lines are $G_+$ and $G_-$ curves, respectively. The green dotted line is $%
E_{0}^{pole}$ and the black dotted lines are $E_{n}^{pole}$. The data by
numerics are indicated by open circles, which agree excellently with the
zeros of the $G$-functions.}
\label{G-function}
\end{figure}

\begin{figure}[tbp]
\includegraphics[width=\linewidth]{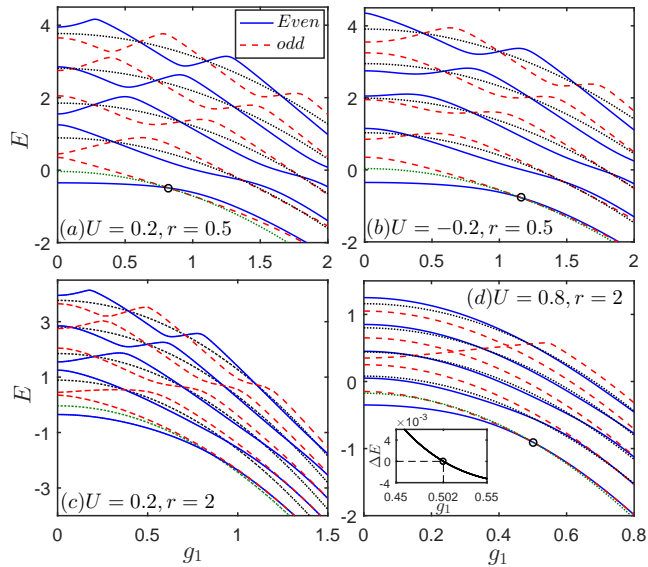}
\caption{ (Color online) The spectra for the anisotropic RSM at $\Delta=0.7$%
. The green dotted line is $E_{0}^{pole}$ by Eq. (\protect\ref{pole0}) and
the black dotted line is $E_{n}^{pole}$ by Eq. (\protect\ref{pole_m}). The
crossing points of the ground-state and the first-excited state are marked
by open circles, which are at  $g_{1c}= 0.82$, $E = -0.49$
in panel (a), $1.159$, $-0.77$ in panel (b), and $0.502$, $-0.91$ in panel (d). The inset in panel (d) shows  the energy difference  $\Delta E$ between the two lowest energy
levels.}
\label{spectrum}
\end{figure}

The zeros of this $G$-function can give the regular spectrum. The
eigenfunction is then obtained through Eq. (\ref{wave1}) with the
eigenenergy. To demonstrate this point, we plot the $G$-function for $\Delta
=0.7,g_{1}=0.8,U=0.2,r=0.5$ and $2$ in Fig. \ref{G-function}. The zeros
reproduce all regular spectra, which can be confirmed by the numerical exact
diagonalizations in the truncated Fock space. The spectra for a few typical
values of $U$ and $r$ are displayed in Fig. \ref{spectrum}.

\section{First-order QPTs}

The level crossings in the QRM and its variants are ubiquitous as long as
the parity is conserved. But the level crossings of the ground-state and the
first-excited state energy does not always exist. This special level
crossing is just a criterion of the first-order quantum phase transition,
because the first derivative of the ground-state energy with respect to the
coupling constant is discontinuous.

In the two-level atom and the cavity coupling systems, the well-defined pole
structure of the derived transcendental functions are very useful. To the
best of our knowledge, the characteristics of these poles can be used to
analyze the level distribution~\cite{Braak}, level crossings \cite{Zhong},
and spectra collapse ~\cite{Chen2012, duan}. These subtle issues are however
hardly settled by the numerics as well as the analytical treatments without
poles. We will use the first pole to locate the level crossing points of the
ground-state and the first-excited state in the following.

To show the level crossings in the present ARSM in general, we also analyze
the pole structure of the derived G-function (\ref{G-Func}). The vanishing
coefficient of $f_{m}$\ in Eq. (\ref{relation2}) yields the $m$-th ($m>0$)
pole of the G-functions
\begin{equation}
E_{m}^{pole}=\left( 1-U^{2}\right) m-\lambda _{+}-\frac{U\Delta }{2}.
\label{pole_m}
\end{equation}%
If the right-hand-side of Eq. (\ref{relation2}) is also zero, we can then
obtain the values of $g_1$ and $E$ at crossing point above the first pole if
$\Delta, U$, and $r$ are given. All crossing points above the first pole
line shown in Fig. \ref{spectrum} are consistent with the analytical
predictions.

In particular, because of $f_0=1$ in the present scheme, the first pole is
however given by the vanishing denominator of $e_{0}$ in Eq. (\ref{relation1}%
) for $m=0$
\begin{equation}
E_{0}^{pole}=-\frac{U\Delta }{2+2\sqrt{1-U^{2}}}-\frac{\lambda _{-}\left( 1-%
\sqrt{1-U^{2}}\right) /U+\lambda _{+}}{\sqrt{1-U^{2}}}.  \label{pole0}
\end{equation}%
which is not included in the general poles described in Eq. (\ref{pole_m}).
This first pole equation is exactly reduced to that in the isotropic RSM ~%
\cite{Xie} if setting $r=1$, and the anisotropic QRM ~\cite{Fanheng} if $U=0$%
.

The poles given in Eqs. (\ref{pole0}) and (\ref{pole_m}) are also exhibited
in Fig. \ref{G-function} with dotted lines. The G-curves at these poles
indeed show the diverging behavior. As usual, if both $e_{n}$ and $f_{n}$ in
G-function (\ref{G-Func}) are analytic at these pole energies, one obtains
the Juddian solutions for doubly degenerate states \cite{Judd}. In this
case, two adjacent energy levels with the even and odd parity can
simultaneously intersect with the associated pole line $E_{n}^{pole},
n=0,1,2,...$, in the energy spectra. Below we  focus on the possible
level crossing of the first two lowest levels associated with the first pole
($E_{0}^{pole}$) in this model, and skip the discussions on the whole
Juddian solutions, which are in fact similar to the previous ones in both
the isotropic QRM \cite{Braak} and the anisotropic QRM \cite{Fanheng}.

At the first pole energy $E=E_{0}^{pole}$, the denominator of $e_{0}$ is
zero, so $e_{0}$ is analytic only if its  numerator in Eq. (\ref%
{relation1}) vanishes, yielding a special coupling strength where the
ground-state energy and the first-excited state energy cross, i. e. the
critical point of the first-order QPT
\begin{equation}
g_{1c}=\sqrt{\frac{\Delta \left( 1-U^{2}\right) }{U\left( 1+r^{2}\right)
+1-r^{2}}}.  \label{1st}
\end{equation}%
It can be reduced to that in the anisotropic QRM if setting $U=0$~\cite%
{Fanheng}, and in the isotropic RSM if $r=1$ ~\cite{Xie}. The critical
points given by Eq. (\ref{1st}) are demonstrated in Fig. \ref{spectrum} (a),
(b), and (d) with open circles. The absence of the first-order QPT in Fig. %
\ref{spectrum} (c) is due to the fact that no real solution exists in Eq. (%
\ref{1st}) at those parameters. Especially in Fig. 2(d), the two lowest
levels are too close to be discerned after the crossing point. Fortunately,
it can be detected by the present analytical study.

One can find that in the present ARSM, the first-order QPT can be induced by
the presence of either the anisotropy or the nonlinear Stark coupling on
the condition that $r<\sqrt{\frac{1+U}{1-U}}$. The first-order QPT is still
possible in this model even for $r>1$, which is however forbidden in the
anisotropic QRM. In addition, the first-order QPT can also occurs when $U<0$
if $r<1$, however it is impossible in the isotropic RSM for $U<0$ ~\cite{Xie}%
. The parameter range for the occurrence of the first-order QPT in the
present model is much more broader than the previous ones.

The first-order phase transition occurs at finite model parameters only if
the Juddian solution associated with the first pole (\ref{pole0}) exists. In
the isotropic QRM ($U=0,g_{1}=g_{2}$), the first Juddian solution ($n=0$) is
absent \cite{Braak}. Only if $U\neq 0$ and/or $g_{1}\neq g_{2}$, such a
Juddian solution would appear. For special values of the model parameters,
the first pole can be lifted because both the numerator and the denominator
of $e_{0}$ vanishes. Thus $G_{\pm }\left( E\right) \neq 0$ in this case, the
eigenvalues therefore have no definite parity, and a double degeneracy of
the eigenvalues occurs.

As found in Ref. ~\cite{Xie}, the first crossing energy is $-\Delta /(2U)$
in the isotropic RSM. If $U$ is absent, the crossing energy is negatively
infinite, which cannot be reached by the first two levels, consistent with
the absence of the first-order QPT in the isotropic QRM. But for finite $U$,
the crossing energy becomes finite, so it is possible that the first two
levels cross somewhere at this energy. This possibility is just induced by the
the Stark coupling. For the anisotropic QRM, in the extremely case, e.g.
RWA, there is always an eigenenergy $-\Delta /2$, the adjacent energy level
must cross this energy as the coupling strength increases. With the addition of the counter rotating-wave terms, as long as its coupling strength is weaker than that of the rotating-wave term (i.e., $r<1$), the level crossing of the first two levels must happen~\cite{Fanheng}. The first-order QPT   however disappears if $r>1$, which obviously  can be
attributed to the competition between the rotating and counter-rotating
interaction terms.  In the present
more complicated ARSM, both the Stark coupling and the anisotropy cooperate
to enlarge the parameter range of the first-order QPT, as demonstrated
above. In short, the first-order QPT depends on the interplay among the
Stark coupling, the rotating-wave coupling, and the counter-rotating wave
coupling.

\section{Continuous quantum phase transitions at $U=\pm 1$}

It is known that the continuous QPT would occur in the QRM in the infinite
frequencies ratio $\Delta/\omega $ ~\cite{plenio}, which should also happen
in the present ARSM. In this section, we do not discuss this obvious
continuous QPT, but explore the continuous QPT in the ARSM at finite
frequencies ratio $\Delta/\omega $ for the special Stark coupling $%
\left\vert U\right\vert =1$.

Note that the solution at $U=\pm 1$ cannot be given in the above BOA, but
can be obtained in another way as described in Ref. \cite{Xie}. To this end,
we may write the ARSM Hamiltonian in the basis of $\sigma_{x}$
\begin{eqnarray}
H&=&\left( \frac{\Delta }{2}+Ua^{\dagger }a\right) \sigma_{z}+a^{\dagger
}a+\alpha \left( a^{\dagger }+a\right) \sigma _{x}  \notag \\
&&+\kappa \alpha \left( a-a^{\dagger }\right) i\sigma _{y},  \label{H2}
\end{eqnarray}%
Comparing with the Hamiltonian (\ref{Hamiltonian}), we have $\alpha =({%
g_{1}+g_{2}})/2,\kappa \alpha =({g_{1}-g_{2}})/2$, where $\kappa
=(1-r)/(1+r)\leqslant 1$. If $g_{1}=g_{2}$, i.e., $\kappa =0,$ the isotopic
RSM is recovered, while if $g_{2}=0$, i.e., $\kappa =1$, it corresponds to
the RSM in the RWA.

We can map the ARSM Hamiltonian at $U=1$ to an effective quantum oscillator,
the details are given in Appendix B. The solutions to the eigenenergies are
given by solving Eq. (\ref{OS_energy}) self-consistently. Obviously, the
whole energy spectra separates into two branches: the upper one $E>-\frac{%
\Delta }{2}-2\kappa ^{2}\alpha ^{2}$ and the lower one $E<-\frac{\Delta }{2}%
-2\alpha ^{2}$. The real lower spectra only exist before the critical
coupling $\alpha _{c}^{+}$
\begin{equation}
\alpha _{c}^{+}=\sqrt{\frac{1-\Delta +\kappa }{2}},  \label{critical_U1}
\end{equation}%
and the upper bound of the low spectra is $E_{c}^{+}=-\frac{\Delta }{2}%
-2\alpha ^{2}$.

For $U=-1$, all results can be straightforwardly obtained by replacing $%
\Delta $ and $\kappa $ by $-\Delta $ and $-\kappa $, respectively. The
corresponding critical coupling strength
\begin{equation}
\alpha _{c}^{-} =\sqrt{\frac{1+\Delta -\kappa }{2}},  \label{critical_Um1}
\end{equation}
and the upper bound of the lower energy spectra  is $E_{c}^{-} =\frac{%
\Delta }{2}-2\alpha ^{2}$.

\begin{figure}[tbp]
\includegraphics[width=\linewidth]{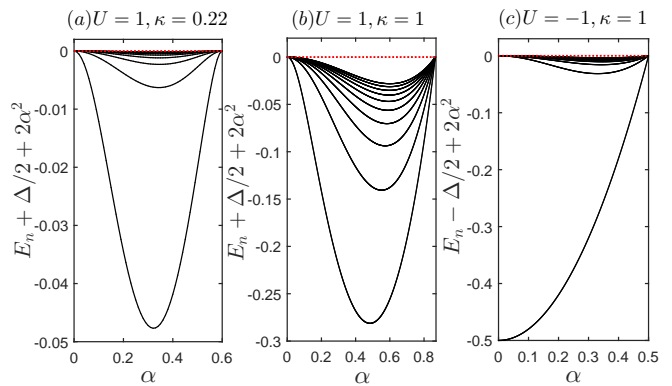}
\caption{ (Color online) The differences of the  several low energy levels
and $E_{c}^{+}$ (left, middle), $E_{c}^{-}$ (right) as a function of $%
\protect\alpha$ by solving Eq. (\protect\ref{OS_energy}) self-consistently for $
\Delta=0.5$. The dotted red horizontal lines correspond to the upper bounds.}
\label{Energy_collapse}
\end{figure}

Figure \ref{Energy_collapse} presents  several low energy levels in the
lower spectra with different parameters. To show more clear, the energy is
shifted by the corresponding upper bound of the low energy spectra. All
levels close at the critical points given by Eq. (\ref{critical_U1}) for $%
U=1 $ or Eq. (\ref{critical_Um1}) for $U=-1$.

\begin{figure}[tbp]
\includegraphics[width=\linewidth]{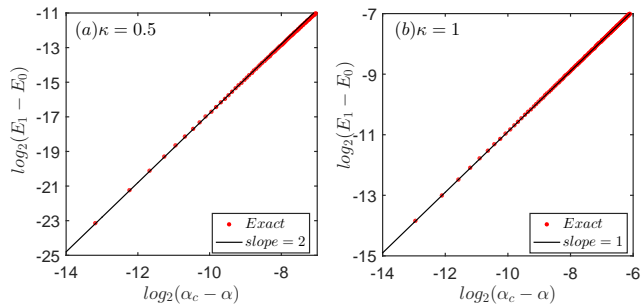}
\caption{(Color online) The log-log plot of energy gap $E_1-E_0$ as a
function of $\protect\alpha_c-\protect\alpha$ at $U=1,\Delta=0.5$ for $%
\protect\kappa =1 $, i.e. RWA (right) and $\protect\kappa =0.5$, i.e.
non-RWA (left). }
\label{Energy_gap_exponent}
\end{figure}

\textsl{Energy gap:} The low energy spectra equation (\ref{OS_energy}) for $%
U=1$ can be rewritten as
\begin{eqnarray}
\frac{2b\alpha ^{2}}{\sqrt{x}}+\left( b-\kappa -2\alpha ^{2}-x\right) \sqrt{x%
}  \notag \\
=\sqrt{x+2\alpha ^{2}\left( 1-\kappa ^{2}\right) }\left( 2n+1\right) ,
\label{low_limit}
\end{eqnarray}%
where $x=E_{c}^{+}-E$ and $b=2\left( \alpha _{c}^{+}\right) ^{2}-2\alpha
^{2} $. Below we omit the superscript $+$ in $\alpha _{c}^{+}$ and $E_{c}^{+}$
for simplicity. When $\alpha \rightarrow \alpha _{c},E\rightarrow E_{c}$, $%
x,b\rightarrow 0$. Note that the right-hand side of Eq. (\ref{low_limit}) is
finite in this limit, the first term of the left-hand side reveals that $x$
must be of the following form:
\begin{equation}
x=rb^{2}+O(b^{3}),
\end{equation}
or else the left-hand-side is infinite. So the energy gap between the ground
state (n=0) and the first excited state (n=1) is
\begin{equation}
E_{g}=E_{1}-E_{0}\propto \left\vert \alpha -\alpha _{c}\right\vert ^{2}.
\end{equation}%
This is to say that, at $U=1$, the energy gap closes at $\alpha _{c}$ with a
critical exponent $2$. Generally, in the continuous QPT, the energy gap
displays a universal scaling behavior, $E_{g}\propto \left\vert
\alpha-\alpha_{c}\right\vert ^{z\nu}$, where $z(\nu)$ is the (dynamics)
critical exponent ~\cite{Sachdev}. This gap exponent can be confirmed by
solving Eq. (\ref{OS_energy}) self-consistently, as shown in the left plot
of Fig. \ref{Energy_gap_exponent}.

If $\kappa =1$, the counter-rotating wave coupling is absent, this is just
the RWA, and Eq. (\ref{low_limit}) then becomes
\begin{equation}
\frac{2b\alpha ^{2}}{x}+\left( b-1-2\alpha ^{2}-x\right) =\left( 2n+1\right).
\end{equation}%
In this case, $x$ must be of the form
\begin{equation}
x=rb+O(b^{2}),
\end{equation}%
so the energy gap is
\begin{equation}
E_{g}\propto \left\vert \alpha -\alpha _{c}\right\vert ,
\end{equation}%
with the gap exponent $z\nu=1$, which is also verified in the right plot of
Fig. \ref{Energy_gap_exponent}.

Very interestingly, the real solution for the eigenenergy can even exist
above $\alpha _{c}$ in the RWA, in sharp contrast to any $\left\vert \kappa
\right\vert <1$ case. Equation (\ref{OS_energy}) at $\kappa =1$ gives

\begin{equation}
E_{n}^{\pm }=n\pm \sqrt{n^{2}+\left( \frac{\Delta }{2}+2n\right) \frac{%
\Delta }{2}+4\alpha ^{2}\left( n+1\right) },  \label{RWAU1}
\end{equation}%
where $+(-)$ denotes the upper (lower) spectra. The extension to the $U=-1$
is straightforward, and will not be presented here.

The derivative of the energy level with respect to $n$ in the lower spectra
is given by%
\begin{equation}
\frac{dE_{n}^{-}}{dn}=1-\frac{2\left( n+\frac{\Delta }{2}\right) +4\alpha
^{2}}{2\sqrt{\left( n+\frac{\Delta }{2}\right) ^{2}+4\left( n+1\right)
\alpha ^{2}}},  \label{RWA_U1_low}
\end{equation}%
the extremal condition happens exactly at the critical coupling  by Eq.
(\ref{critical_U1}) with $\kappa=1$
\begin{equation*}
\alpha _{c}=\sqrt{1-\frac{\Delta }{2}}.
\end{equation*}%
It is obvious that, for $\alpha <\alpha _{c}$, the low energy spectra
increase with $n$, while for $\alpha >\alpha _{c}$, the low energy spectra
decrease with $n$. Thus the ground-state energy corresponds to $n=0$ for $%
\alpha <\alpha _{c}$. However, for $\alpha >\alpha _{c}$, the ground-state
energy is surprisingly corresponding to the infinite $n$,
\begin{equation}
E_{0}=E_{n\rightarrow \infty }^{-}=-\frac{\Delta }{2}-2\alpha ^{2}.
\label{limit2}
\end{equation}%
So the gap between the first-excited state and ground state always vanishes
for $\alpha >\alpha _{c}$, because
\begin{equation*}
E_{g}=\lim_{n\rightarrow \infty }\left( E_{n-1}^{-}-E_{n}^{-}\right) =0.
\end{equation*}%
It just demonstrates the appearance of photonic Goldstone modes above a
critical point. In the Dicke model with infinite two-level atoms in the RWA,
a Goldstone soft mode appears above a critical point as a consequence of the
$U(1)$ symmetry breaking ~\cite{Ciuti}. Here although the RWA is also made,
 only one two-level atom is involved. In the ARSM under the RWA, the
system also possesses $U(1)$ symmetry, and is broken above the critical
points.

In the ground state of the RSM under the RWA, the photonic number $n=0$
below $\alpha _{c}$, but $n\rightarrow \infty $ above $\alpha _{c}$,
suggesting a special superradiant phase.

\section{Conclusion}

In this work, we find that both the first-order and the continuous QPTs are
present in the ARSM where both the nonlinear Stark coupling and the
anisotropic dipole linear coupling are present. Among the previous QRM and
the Dicke model, as well as their generalized models, both types of QPTs
have not been observed in the same model in the literature.

The first-order QPT is detected analytically by the pole structure of
G-functions based on the BOA. The critical coupling strength of the phase
transitions is obtained analytically, which is determined by both anisotropy
and the nonlinear Stark coupling. On the other hand, the continuous QPT is
also found in this model at the special values of the Stark coupling
strength $U=\pm 1$ for the closing energy spectra at the critical points.
The energy gap follows an universal power-law scaling ansatz $E_{g}\propto
\left\vert \alpha-\alpha_{c}\right\vert ^{z\nu}$ in any ARSMs. The energy-gap exponent $z\nu =1$ for $\kappa =1$, i.e., the RWA; while $2$ for $\kappa <1$,
indicating that the presence of any counter-rotating wave terms would change
the universality class of this model. In the RWA, since the gap is always
closed above the critical points, one phase having Goldstone mode gapless
excitations then appears.

The continuous QPT undergoes in the QRM at the infinite frequency ratio,
occurs in the Dicke model in the thermodynamic limit. These prerequisite
conditions are however not required in the present ARSM for the occurrence
of the QPTs. Especially in the RWA, the critical coupling can be weak
sufficiently by tuning the qubit frequency in the ARSM. We therefore believe
that the continuous QPTs might be easily demonstrated experimentally or in
quantum simulations based on some solid-state devices, such as the cavity
(circuit) QEDs and the ion-trap, where the ARSM Hamiltonian can be realized.

\textbf{ACKNOWLEDGEMENTS} We acknowledge useful discussions with Lei Cong. This work is supported by the National Science
Foundation of China (Nos. 11674285, 11834005), the National Key Research and
Development Program of China (No. 2017YFA0303002),

$^{*}$ Email:qhchen@zju.edu.cn


\begin{appendix}

\section{Derivation of G-function of the ARSM by BOA}

In this Appendix, we derive a transcendant function to the ARSM by BOA.
By using the transformation (\ref{P1}) we obtain the transformed
Hamiltonian in the matrix form
\begin{widetext}
\begin{equation}
H_{1}=PHP^{-1}=\left(
\begin{array}{ll}
a^{\dagger }a+\beta \left( a+a^{\dagger }\right) +\left( \frac{\lambda _{+}}{%
\beta }-\beta \right) a^{\dagger } & \;\;\;\;-\left( \frac{1}{2}\Delta
+Ua^{\dagger }a\right) -\frac{\lambda _{-}}{\beta }a^{\dagger } \\
\;\;\;\;-\left( \frac{1}{2}\Delta +Ua^{\dagger }a\right) +\frac{\lambda _{-}%
}{\beta }a^{\dagger } & \;a^{\dagger }a-\beta \left( a+a^{\dagger }\right)
-\left( \frac{\lambda _{+}}{\beta }-\beta \right) a^{\dagger }%
\end{array}%
\right) ,
\end{equation}%
\end{widetext}where $\lambda _{\pm }=\left( g_{1}^{2}\pm g_{2}^{2}\right) /2$
and$\;\beta =\sqrt{g_{1}g_{2}}$. It can be expressed in terms of new
operator $A$ defined in Eq. (\ref{dis}) with the four matrix elements below
\begin{eqnarray*}
H_{11} &=&A^{\dagger }A-\left( w-\beta \right) \left( A^{\dagger }+A\right)
\\
&&+w^{2}-2\beta w+\left( \frac{\lambda _{+}}{\beta }-\beta \right) \left(
A^{\dagger }-w\right) ,
\end{eqnarray*}%
\begin{eqnarray*}
H_{22} &=&A^{\dagger }A-\left( w+\beta \right) \left( A^{\dagger }+A\right)
\\
&&+w^{2}+2\beta w-\left( \frac{\lambda _{+}}{\beta }-\beta \right) \left(
A^{\dagger }-w\right) ,
\end{eqnarray*}%
\begin{eqnarray*}
H_{12} &=&-\frac{1}{2}\Delta -U\left[ A^{\dagger }A-w\left( A^{\dagger
}+A\right) +w^{2}\right]  \\
&&-\frac{\lambda _{-}}{\beta }\left( A^{\dagger }-w\right) ,
\end{eqnarray*}%
\begin{eqnarray*}
H_{21} &=&-\frac{1}{2}\Delta -U\left[ A^{\dagger }A-w\left( A^{\dagger
}+A\right) +w^{2}\right]  \\
&&+\frac{\lambda _{-}}{\beta }\left( A^{\dagger }-w\right) .
\end{eqnarray*}

In terms of the  eigenfunction (\ref{wave1}), we can obtain the Schr\"{o}dinger
equations for both upper and lower levels, then projecting both sides of the
Schr\"{o}dinger equations onto $_{A}\left\langle m\right\vert \ $ gives
\begin{eqnarray}
&&\left[ \Gamma _{m}-\left( \frac{\lambda _{+}}{\beta }+\beta \right) w-E%
\right] e_{m}+\left( \frac{\lambda _{+}}{\beta }-\beta \right) e_{m-1}
\notag \\
&&+\left[ -\frac{1}{2}\Delta +\frac{\lambda _{-}}{\beta }w-U\Gamma _{m}%
\right] f_{m}-\frac{\lambda _{-}}{\beta }f_{m-1}  \notag \\
&&-\left( w-\beta \right) \Lambda _{m}+Uw\digamma _{m}  \notag \\
&=&0,  \label{ARS_S1}
\end{eqnarray}%
\begin{eqnarray}
&&\left[ -\frac{1}{2}\Delta -U\Gamma _{m}-\frac{\lambda _{-}}{\beta }w\right]
e_{m}+\frac{\lambda _{-}}{\beta }e_{m-1}  \notag \\
&&+\left[ \Gamma _{m}+\left( \frac{\lambda _{+}}{\beta }+\beta \right) w-E%
\right] f_{m}-\left( \frac{\lambda _{+}}{\beta }-\beta \right) f_{m-1}
\notag \\
&&+Uw\Lambda _{m}-\left( w+\beta \right) \digamma _{m}  \notag \\
&=&0,  \label{ARS_S2}
\end{eqnarray}%
where
\begin{eqnarray*}
\Lambda _{m} &=&(m+1)e_{m+1}+e_{m-1}, \\
\quad \digamma _{m} &=&(m+1)f_{m+1}+f_{m-1}, \\
\quad \Gamma _{m} &=& m+w^{2}.
\end{eqnarray*}%
Multiplying the Eq. (\ref{ARS_S1}) by $\left( w+\beta \right) $ and Eq. (\ref%
{ARS_S2}) by $Uw$, we have
\begin{eqnarray}
&&\left( w+\beta \right) \left[ \Gamma _{m}-\left( \frac{\lambda _{+}}{\beta
}+\beta \right) w-E\right] e_{m}  \notag \\
&&-\left( w+\beta \right) \left( w-\beta \right) \Lambda _{m}+\left( w+\beta
\right) \left( \frac{\lambda _{+}}{\beta }-\beta \right) e_{m-1}  \notag \\
&&+\left( w+\beta \right) \left[ -\frac{1}{2}\Delta +\frac{\lambda _{-}}{%
\beta }w-U\Gamma _{m}\right] f_{m}  \notag \\
&&+\left( w+\beta \right) Uw\digamma _{m}-\left( w+\beta \right) \frac{%
\lambda _{-}}{\beta }f_{m-1}  \notag \\
&=&0  \label{ARS_S1new}
\end{eqnarray}%
\begin{eqnarray}
&&Uw\left[ -\frac{1}{2}\Delta -U\Gamma _{m}-\frac{\lambda _{-}}{\beta }w%
\right] e_{m}  \notag \\
&&+\left( Uw\right) ^{2}\Lambda _{m}+Uw\frac{\lambda _{-}}{\beta }e_{m-1}
\notag \\
&&+Uw\left[ \Gamma _{m}+\left( \frac{\lambda _{+}}{\beta }+\beta \right) w-E%
\right] f_{m}  \notag \\
&&-Uw\left( w+\beta \right) \digamma _{m}-Uw\left( \frac{\lambda _{+}}{\beta
}-\beta \right) f_{m-1}  \notag \\
&=&0.  \label{ARS_S2new}
\end{eqnarray}%
Summation of  Eq. (\ref{ARS_S1new}) and Eq. (\ref{ARS_S2new})  gives
\begin{eqnarray}
&&\left(
\begin{array}{c}
\left( w+\beta \right) \left[ \Gamma _{m}-\left( \frac{\lambda _{+}}{\beta }%
+\beta \right) w-E\right]  \\
+Uw\left[ -\frac{1}{2}\Delta -U\Gamma _{m}-\frac{\lambda _{-}}{\beta }w%
\right]
\end{array}%
\right) e_{m}  \notag \\
&&+\left[ \left( Uw\right) ^{2}-\left( w+\beta \right) \left( w-\beta
\right) \right] \Lambda _{m}  \notag \\
&&+\left[ \left( w+\beta \right) \left( \frac{\lambda _{+}}{\beta }-\beta
\right) +Uw\frac{\lambda _{-}}{\beta }\right] e_{m-1}  \notag \\
&&-\left[ \left( w+\beta \right) \frac{\lambda _{-}}{\beta }+Uw\left( \frac{%
\lambda _{+}}{\beta }-\beta \right) \right] f_{m-1}  \notag \\
&&+\left(
\begin{array}{c}
\left( w+\beta \right) \left[ -\frac{1}{2}\Delta +\frac{\lambda _{-}}{\beta }%
w-U\Gamma _{m}\right]  \\
+Uw\left[ \Gamma _{m}+\left( \frac{\lambda _{+}}{\beta }+\beta \right) w-E%
\right]
\end{array}%
\right) f_{m}  \notag \\
&=&0.  \label{simple}
\end{eqnarray}%
To remove the term containing $\Lambda _{m}$, the displacement should be%
\begin{equation}
w=\frac{\beta }{\sqrt{1-U^{2}}}.  \label{shift_ARS}
\end{equation}%
Then by Eq. (\ref{simple}) we have $\allowbreak $\
\begin{widetext}
\begin{equation}
e_{m}=\frac{\left\{ \frac{1}{2}\Delta -\frac{\lambda _{-}}{\beta }w+U\Gamma
_{m}-\frac{Uw}{\left( w+\beta \right) }\left[ \Gamma _{m}+\frac{\lambda
_{+}+\beta ^{2}}{\beta }w-E\right] \right\} f_{m}-\left[ \frac{\lambda
_{+}-\beta ^{2}}{\beta }+\frac{Uw\lambda _{-}}{\left( w+\beta \right) \beta }%
\right] e_{m-1}+\left[ \frac{\lambda _{-}}{\beta }+\frac{Uw\left( \lambda
_{+}-\beta ^{2}\right) }{\beta \left( w+\beta \right) }\right] f_{m-1}}{%
\Gamma _{m}-\frac{\lambda _{+}+\beta ^{2}}{\beta }w-E-\frac{Uw}{\left(
w+\beta \right) }\left[ \frac{1}{2}\Delta +U\Gamma _{m}+\frac{\lambda _{-}}{%
\beta }w\right] }.  \label{relation1}
\end{equation}%
Inserting Eq. (\ref{relation1}) to  Eq. (\ref{ARS_S1}) at $m-1$ gives
\begin{eqnarray}
&&\left( \frac{Uw}{w-\beta }-\frac{\frac{1}{2}\Delta -\frac{\lambda _{-}}{%
\beta }w+\frac{U\beta }{w+\beta }\Gamma _{m}-\frac{Uw}{\left( w+\beta
\right) }\left[ \frac{\lambda _{+}+\beta ^{2}}{\beta }w-E\right] }{\Theta (E)%
}\right) f_{m}=-\frac{Uw-\frac{\lambda _{-}}{\beta }}{m\left( w-\beta
\right) }f_{m-2}-\frac{\frac{\lambda _{+}}{\beta }-w}{m\left( w-\beta
\right) }e_{m-2}  \notag \\
&&-\left( \frac{\frac{\lambda _{-}}{\beta }+\frac{Uw\left( \lambda _{+}-\beta
^{2}\right) }{\beta \left( w+\beta \right) }}{\Theta (E)}+\frac{\frac{1}{2}%
\Delta -\frac{\lambda _{-}}{\beta }w+U\Gamma _{m-1}}{m\left( w-\beta \right)
}\right) f_{m-1}-\left( \frac{\frac{\lambda _{+}-\beta ^{2}}{\beta }+\frac{%
Uw\lambda _{-}}{\left( w+\beta \right) \beta }}{\Theta (E)}+\frac{\Gamma
_{m-1}-\frac{\lambda _{+}+\beta ^{2}}{\beta }w-E}{m\left( w-\beta \right) }%
\right) e_{m-1},  \label{relation2}
\end{eqnarray}%
where%
\begin{equation*}
\Theta (E)=\sqrt{1-U^{2}}\Gamma _{m}-\frac{\lambda _{+}+\beta ^{2}}{\beta }%
w-E-\frac{Uw}{\left( w+\beta \right) }\left( \frac{\Delta }{2}+\frac{\lambda
_{-}}{\beta }w\right) .
\end{equation*}%

\end{widetext}Starting from $f_{0}=1,e_{-1}=f_{-1}=0$, then $e_{0}$ can be
obtained by Eq. (\ref{relation1}) at $m=0$ and $f_{1}$ by Eq. (\ref%
{relation2}). By a similar procedure, we can obtain any $m$th coefficients
$e_{m}$ and $f_{m}$.

Alternatively, the eigenfunction can also be expanded in the another Bogoliubov operator $B$ with the opposite displacement as
\begin{equation}
\left\vert {}\right\rangle _{B}=\left( \
\begin{array}{l}
\sum_{n=0}^{\infty }(-1)^{n}\sqrt{n!}f_{n}\left\vert n\right\rangle _{B} \\
\sum_{n=0}^{\infty }(-1)^{n}\sqrt{n!}e_{n}\left\vert n\right\rangle _{B}%
\end{array}%
\right) ,  \label{wave2}
\end{equation}%
due to the parity symmetry. Here $\left\vert n\right\rangle _{B}$ is defined\ in
\ the similar way as $\left\vert n\right\rangle _{A}$.

Assuming both wavefunctions (\ref{wave1}) and (\ref{wave2}) are the true
eigenfunction for a nondegenerate eigenstate with eigenvalue $E$, they
should be proportional to  each other, \textsl{i.e.,} $\left\vert
{}\right\rangle _{A}=r\left\vert {}\right\rangle _{B}$, where $r$ is a
complex constant. Projecting both sides of this identity onto the original
vacuum state $_{a}\left\langle 0\right\vert $, we have
\begin{eqnarray*}
\sum_{n=0}^{\infty }\sqrt{n!}e_{n}~_{a}\langle 0|n\rangle _{A}
&=&r\sum_{n=0}^{\infty }\sqrt{n!}(-1)^{n}f_{n}~_{a}\langle 0|n\rangle _{B},
\\
\sum_{n=0}^{\infty }\sqrt{n!}f_{n}~_{a}\langle 0|n\rangle _{A}
&=&r\sum_{n=0}^{\infty }\sqrt{n!}(-1)^{n}e_{n}~_{a}\langle 0|n\rangle _{B},
\end{eqnarray*}%
where
\begin{equation*}
\sqrt{n!}~_{a}{\langle }0|n{\rangle }_{A}=(-1)^{n}\sqrt{n!}~_{a}{\langle }0|n%
{\rangle }_{B}=e^{-w^{2}/2}w^{n}.
\end{equation*}%
Eliminating the ratio constant $r$ gives
\begin{equation*}
\left( \sum_{n=0}^{\infty }e_{n}w^{n}\right) ^{2}=\left( \sum_{n=0}^{\infty
}f_{n}w^{n}\right) ^{2}.
\end{equation*}%
Immediately, we obtain the following well-defined transcendental function,
the s-ocalled $G$-function, as
\begin{equation}
G_{\mp }\left( E\right) =\sum_{n=0}^{\infty }\left( e_{n}\pm f_{n}\right)
w^{n}=0,
\end{equation}%
where $\mp $ corresponds to odd(even) parity.   Interestingly,  this $G$-function can be reduced
to those of the RSM if  $r=1$~\cite{Xie},   the anisotropic  QRM if $U=0$ ~\cite{Fanheng}, and the isotropic QRM if $U=0$ and $r=1$~\cite{Braak}.  It is worth noting that even in the presence of both the nonlinear Stark coupling  and the anisotropic linear dipole coupling in this generalized model,  the G-function can still be obtained within the BOA in the concise way.

\section{Solutions to the Anisotropic Rabi-Stark Model at $U=1$}

In this Appendix, we turn to the ARSM at $U=1$ which solutions cannot be
covered in Appendix A.

In terms of the position and momentum representations, $x=\frac{1}{\sqrt{2}}%
\left( a^{\dagger }+a\right) ,p=\frac{i}{\sqrt{2}}\left( a^{\dagger
}-a\right) $, Hamiltonian (\ref{H2}) at $U=1$ can be written as
\begin{equation}
H_{2}=\left(
\begin{array}{ll}
p^{2}+x^{2}-1+\frac{\Delta }{2} & \alpha \sqrt{2}x+i\kappa \alpha \sqrt{2}p
\\
\alpha \sqrt{2}x-i\kappa \alpha \sqrt{2}p & \;-\frac{\Delta }{2}%
\end{array}%
\right) .
\end{equation}%
For the eigenfunction $\Psi =\left( \phi _{1},\phi _{2}\right) ^{T}$, the
Schr\"{o}dinger equations for the upper and lower level now \ are
\begin{eqnarray*}
\left( p^{2}+x^{2}-1+\frac{\Delta }{2}\right) \phi _{1}+\left( \alpha \sqrt{2%
}x+i\kappa \alpha \sqrt{2}p\right) \phi _{2} &=&E\phi _{1}, \\
\left( \alpha \sqrt{2}x-i\kappa \alpha \sqrt{2}p\right) \phi _{1}-\frac{%
\Delta }{2}\phi _{2} &=&E\phi _{2},
\end{eqnarray*}%
where $E$ is the eigenvalue. Inserting $\ \phi _{2}=\frac{\left( \alpha
\sqrt{2}x-i\kappa \alpha \sqrt{2}p\right) }{E+\frac{\Delta }{2}}\phi _{1}$
to the first equation results in the effective one-body Hamiltonian for $%
\phi _{1},$
\begin{equation*}
H_{eff}\phi _{1}=\left( E+1-\frac{\Delta }{2}-\frac{2\kappa \alpha ^{2}}{E+%
\frac{\Delta }{2}}\right) \phi _{1},
\end{equation*}%
where
\begin{equation}
H_{eff}=2\left( 1+\frac{2\kappa ^{2}\alpha ^{2}}{E+\frac{\Delta }{2}}\right) %
\left[ \frac{p^{2}}{2}+\frac{1}{2}\omega _{eff}^{2}x^{2}\right] ,
\label{H_eff}
\end{equation}%
which is just a quantum harmonic oscillator with an effective oscillator
frequency
\begin{equation*}
\omega _{eff}=\sqrt{\frac{1+\frac{2\alpha ^{2}}{E+\frac{\Delta }{2}}}{1+%
\frac{2\kappa ^{2}\alpha ^{2}}{E+\frac{\Delta }{2}}}}.
\end{equation*}%
So the eigenenergy is then expressed as
\begin{eqnarray}
&&\frac{\left( E+1-\frac{\Delta }{2}\right) \left( E+\frac{\Delta }{2}%
\right) -2\kappa \alpha ^{2}}{E+\frac{\Delta }{2}+2\kappa ^{2}\alpha ^{2}}
\notag \\
&=&\left( 2n+1\right) \sqrt{\frac{E+\frac{\Delta }{2}+2\alpha ^{2}}{E+\frac{%
\Delta }{2}+2\kappa ^{2}\alpha ^{2}}},n=0,1,2,...  \label{OS_energy}
\end{eqnarray}%
Solving this equation self-consistently would give solutions to the ARSM at $%
U=1$.

\end{appendix}


\end{document}